\documentclass{appolb}
\usepackage{epsfig,graphicx}
\begin{document}

\title{Kaon HBT radii from perfect fluid dynamics using the Buda-Lund model}
\author{M. Csan\'ad
\address{E\"otv\"os University, H-1117 Budapest, P\'azm\'any P\'eter s.
1/A, Hungary} \and T. Cs\"org\H{o}
\address{MTA KFKI RMKI, H-1525 Budapest 114, P.O.Box 49, Hungary}}
\maketitle

\begin{abstract}
In this paper we summarize the ellipsoidally symmetric Buda-Lund
model's results on HBT radii. We calculate the Bose-Einstein
correlation function from the model and derive formulas for the
transverse momentum dependence of the correlation radii in the
Bertsch-Pratt system of out, side and longitudinal directions. We
show a comparison to $\sqrt{s_{\rm NN}}=200 GeV$ RHIC PHENIX
two-pion correlation data and make prediction on the same
observable for different particles.
\end{abstract}

\section{Perfect fluid hydrodynamics}
Perfect fluid hydrodynamics is based on local conservation of
entropy and four-momentum. The fluid is perfect if the
four-momentum tensor is diagonal in the local rest frame. The
conservation equations are closed by the equation of state, which
gives the relationship between energy density $\epsilon$, pressure
$p$. Typically $\epsilon - B = \kappa (p+B)$, where $B$ stands for
a bag constant ($B=0$ in the hadronic phase, non-zero in a QGP
phase), and $\kappa$ may be a constant, but can be an arbitrary
temperature dependent function.

There are only a few exact solutions for these equations. One (and
historically the first) is the famous Landau-Khalatnikov solution
discovered more than 50 years
ago~\cite{Landau:1953gs,Khalatnikov:1954aa,Belenkij:1956cd}. This
is a 1+1 dimensional solution, and has realistic properties: it
describes a 1+1 dimensional expansion, does not lack acceleration
and predicts an approximately Gaussian rapidity distribution.

Another renowned solution of relativistic hydrodynamics is the
Hwa-Bjorken solution~\cite{Hwa:1974gn,Chiu:1975hw,Bjorken:1982qr},
which is a simple, explicit and exact, but accelerationless
solution. This solution is boost-invariant in its original form,
but this approximation fails to describe the
data~\cite{Back:2001bq,Bearden:2001qq}. However, the solution
allowed Bjorken to obtain a simple estimate of the initial energy
density reached in high energy reactions from final state hadronic
observables.

There are solutions which interpolate between the above two
solutions~\cite{Csorgo:2006ax,Csorgo:2007ea}, are explicit and
describe a relativistic acceleration.

\section{The Buda-Lund model}

We focus here on the analytic approach in exploring the
consequences of the presence of such perfect fluids in high energy
heavy ion experiments in Au+Au collisions at RHIC. Such exact
analytic solutions were published recently in
refs.~\cite{Csorgo:2006ax,Csorgo:2007ea,Csorgo:2001xm,Csorgo:2003rt,Sinyukov:2004am}.
A tool, that is based on the above listed exact, dynamical hydro
solutions, is the Buda-Lund hydro model of refs.
~\cite{Csorgo:1995bi,Csanad:2003qa}.

The Buda-Lund hydro model successfully describes BRAHMS, PHENIX,
PHOBOS and STAR data on identified single particle spectra and the
transverse mass dependent Bose-Einstein or HBT radii as well as
the pseudorapidity distribution of charged particles in central
Au+Au collisions both at $\sqrt{s_{\rm NN}} = 130$
GeV~\cite{Csanad:2003sz} and at $\sqrt{s_{\rm NN}} = 200$
GeV~\cite{Csanad:2004cj} and in p+p collisions at $\sqrt{s} = 200$
GeV~\cite{Csorgo:2004id}, as well as data from Pb+Pb collisions at
CERN SPS~\cite{Ster:1999ib} and h+p reactions at CERN
SPS~\cite{Csorgo:1999sj,Agababyan:1997wd}. The model is defined
with the help of its emission function; to take into account the
effects of long-lived resonances, it utilizes the core-halo
model~\cite{Csorgo:1994in}. It describes an expanding fireball of
ellipsoidal symmetry (with the time-dependent principal axes of
the ellipsoid being $X$, $Y$ and $Z$).

\section{HBT from the Buda-Lund model}

\begin{figure}
  \includegraphics[width=1\linewidth]{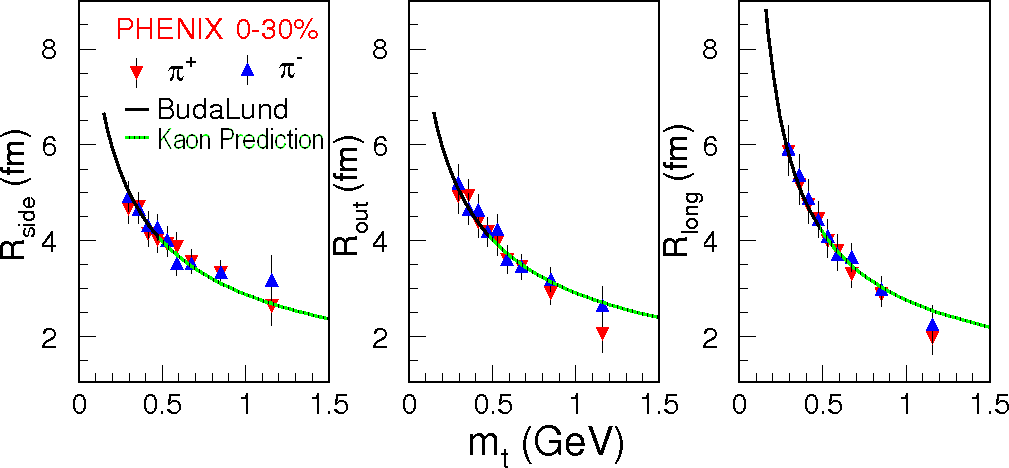}\\
  \caption{HBT radii from the axially Buda-Lund model from
ref.~\cite{Csanad:2004mm}, compared to data of
ref~\cite{Adler:2004rq}. We also show a prediction for kaon HBT
radii on this plot: these overlap with that of pions if plotted
versus transverse mass $m_t$.}\label{f:hbt}
\end{figure}

Let us calculate the two-particle Bose-Einstein correlation
function from the Buda-Lund source function of the Buda-Lund model
as a function of $q=p_1-p_2$, the four-momentum difference of the
two particles. The result is
\begin{equation}
C(q)=1 + \lambda e^{-q_0^2 \Delta \tau_{*}^2 - q_x^2 R_{*,x}^2
-q_y^2 R_{*,y}^2 - q_z^2 R_{*,z}^2}.\label{e:corr}
\end{equation}
with $\lambda$ being the intercept parameter (square of the ratio
of particles emitted from the core versus from the
halo~\cite{Csorgo:1994in}), and
\begin{eqnarray}
\frac{1}{\Delta\tau_*^2} &=& \frac{1}{\Delta \tau^2} + \frac{m_t}{T_0} \frac{d^2}{\tau_0^2},\\
R_{*,x}^2&=&X^2\left(1+m_t(a^2+\dot X^2)/T_0\right)^{-1},\\
R_{*,y}^2&=&Y^2\left(1+m_t(a^2+\dot Y^2)/T_0\right)^{-1},\\
R_{*,z}^2&=&Z^2\left(1+m_t(a^2+\dot Z^2)/T_0\right)^{-1},
\end{eqnarray}
with $\dot X,\dot Y,\dot Z$ being the time-derivative of the
principal axes, $m_t$ the average transverse mass of the pair.
$T_0$ is the central temperature at the freeze-out, $\Delta \tau$
is the mean emission duration and $\tau_0$ is the freeze-out time.
Furthermore, $a$ and $d$ are the spatial and time-like temperature
gradients, defined as $a^2 = \left< \frac{\Delta T}{T}
\right>_\bot$ and $d^2 = \left< \frac{\Delta T}{T} \right>_\tau$.
From the mass-shell constraint one finds $q_0 = \beta_x q_x +
\beta_y q_y + \beta_z q_z$, if expressed by the average velocity
$\beta$. Thus we can rewrite eq.~\ref{e:corr} with modified radii
to
\begin{equation}
C(q)=1 + \lambda_* \exp\left(-\sum_{i,j=x,y,z} R^2_{i,j} q_i
q_j\right)\textnormal{, where}
\end{equation}
\begin{equation}
R_{i,i}^2 = R_{*,i}^2 + \beta_i^2 \Delta \tau_{*}^2\textnormal{, and }
R_{i,j}^2 = \beta_i \beta_j \Delta
\tau_{*}^2,
\end{equation}
From this, we can calculate the radii in the Bertsch-Pratt
frame~\cite{Pratt:1986cc} of out ($o$, pointing towards the
average momentum of the actual pair, rotated from $x$ by an
azimuthal angle $\varphi$), longitudinal ($l$, pointing towards
the beam direction) directions and side ($s$, perpendicular to
both $l$ and $o$) directions. The detailed calculations are
described in ref.~\cite{Csanad:MSc}. These include azimuthally
sensitive oscillating cross-terms. However, due to space
limitations, the angle dependent radii are not shown here. If one
averages on the azimuthal angle, and goes into the LCMS frame
(where $\beta_l = \beta_s = 0$), the Bertsch-Pratt radii are:
\begin{eqnarray}
R_o^2 & = & (R_{*,x}^{-2}+R_{*,y}^{-2})^{-1} + \beta_o^2 \Delta \tau_{*}^2,\\
R_s^2 & = & (R_{*,x}^{-2}+R_{*,y}^{-2})^{-1},\\
R_l^2 & = & R_{*,z}^2.
\end{eqnarray}
These can be fitted then to the data~\cite{Adler:2004rq} as in
ref.~\cite{Csanad:2004mm}, see fig.~\ref{f:hbt}. This allows us to
predict the transverse momentum dependence of the HBT radii of
two-kaon correlations as well: if they are plotted versus $m_t$,
the data of all particles fall on the same curve. This is also
shown for kaons on fig.~\ref{f:hbt}.

\bibliographystyle{prlsty}
\bibliography{Master}

\end{document}